\documentclass[onecolumn,showpacs,amsmath,amssymb,showkeys,pra]{revtex4}

\usepackage{graphicx}
\usepackage{graphics}
\usepackage{epsfig}
\usepackage{amsmath}
\usepackage{float}
\usepackage{multirow}
\usepackage{bigstrut}

\begin{document}

\title{Influence of  Rashba and Dresselhaus spin-orbit
interactions
\\ of
    equal strengths on  electron states in a circular quantum ring  \\in
    the presence of a magnetic field}

\author{V. V. Kudryashov }
 \email{ kudryash@dragon.bas-net.by}

 \author{ A. V. Baran}
\email{ a.baran@dragon.bas-net.by}

\affiliation{Institute of Physics, National Academy of Sciences of
Belarus\\  68 Nezavisimosti Ave., 220072, Minsk, Belarus }

\begin{abstract}
{ Solutions of the Schr\"odinger equation are obtained for an
electron in a two-dimensional
 circular semiconductor quantum ring  in the presence of both  external uniform constant magnetic
 field and  the Rashba and Dresselhaus spin-orbit interactions of equal strengths. Confinement is
 simulated  by a realistic potential well of a finite depth. The dependence of the energy levels
 on a magnetic  field strength,  strength of spin-orbit interaction and a relative ring width
 is presented.}
\end{abstract}



\maketitle

\section{Introduction}

At present it is well established that  motion of an electron in
an inner layer of a semiconductor heterostructure can be treated
as a two-dimensional in $(x,y)$ plane because of the existence of
the confining quantum well along $z$ axis directed perpendicular
to $(x,y)$  plane \cite{val,li}. In connection with the
development of nanotechnology, the study of quantum dots and rings
in heterostructures acquires increasing importance. In
\cite{ban,gro}, a simple but fairly adequate
 model was proposed in which a two-dimensional circular quantum ring corresponds to an axially
 symmetric rectangular potential well

\begin{equation}\label{eq1}
  V_c(\rho) =
  \begin{cases}
    V, &  0<\rho < \rho_i ,\\
    0, &  \rho_i < \rho < \rho_o ,\\
    V, &  \rho_o < \rho < \infty
  \end{cases}
\end{equation}

\noindent of a finite depth $V$, where $\rho = \sqrt {x^2 + y^2}
$, $\rho_i$ and $\rho_o$ are inner and outer radii of a ring.
Notice, that the confining potential of a finite depth was also
used for the description of the quantum dots \cite{kud,chap}.

The influence of the spin-orbit Rashba \cite{ras,byc} and
Dresselhaus \cite{dres} interactions on the electron states in
planar heterostructures are widely studied in recent years. A
uniform constant magnetic field $B$ normal to the plane of the
quantum ring is described by the vector potential ${\bf A}
=\frac{B}{2}(-y,x,0)$. Then the interactions of  Rashba $V_R$ and
Dresselhaus
 $V_D$ are represented by the formulas

\begin{equation}
V_R = \alpha_{\!\scriptscriptstyle R} (\sigma_x P_y - \sigma_y P_x)/\hbar, \quad
 V_D = \alpha_{\!\scriptscriptstyle D} (\sigma_x P_x - \sigma_y P_y )/\hbar,
\end{equation}

\noindent where  ${\bf P} ={\bf p} + q_e {\bf A}$, $q_e$ is the
absolute value of the electron charge, $\sigma_x $ and $\sigma_y $
are standard Pauli spin-matrices. The strengths of these
interactions depend on the used materials. The contribution of two
spin-orbit interactions can be measured with applying various
experimental methods \cite{li, mei}.

In general, the full spin-orbit interaction has the form $V_R +
V_D$ with $\alpha_{\!\scriptscriptstyle R}
\ne\alpha_{\!\scriptscriptstyle D} $. However, the considerable
 attention is spared to the special case \cite{li,schl,bern}, when
 the
spin-orbit interactions of Rashba and Dresselhaus have equal
strength $\alpha_{\!\scriptscriptstyle R} =
\alpha_{\!\scriptscriptstyle D} = \alpha $, which can be
 experimentally achieved due to the fact that the Rashba  interaction strength can be controlled
  by external electric field and the Dresselhaus interaction strength can be varied
   by changing the width of the quantum well along the $z$ axis \cite{val,li}. In the present
    paper for the special case $\alpha_{\!\scriptscriptstyle R} = \alpha_{\!\scriptscriptstyle
D}$, we obtain the wave functions as well as the dependence of the
energy levels on a magnetic field, a spin-orbit interaction
strength and a relative width of the quantum ring which is
described by the potential (\ref{eq1}). The calculations are
performed for the parameter values associated with GaAs.

\section{Exact solution
of the unperturbed Schrodinger equation without the Zeeman
interaction}

The total Hamiltonian of the problem can be written as a sum

\begin{equation}
  H = H_{\scriptscriptstyle 0} + H'
\end{equation}

\noindent of the unperturbed Hamiltonian

\begin{equation}\label{eq4}
  H_{\scriptscriptstyle 0} = \frac{P_x^2 + P_y^2}{2M_{\rm eff}} + V_c (\rho) +
  \frac{\alpha}{\hbar}(\sigma_x - \sigma_y)(P_x + P_y) ,
\end{equation}

\noindent where $M_{\rm eff}$ is the effective electron mass, and
the perturbation describing the Zeeman interaction

\begin{equation}
  H'=\tfrac{1}{2}g \mu_{\!\scriptscriptstyle B} B\sigma_z ,
\end{equation}

\noindent where $g$ is the effective gyromagnetic factor,
$\mu_{\!\scriptscriptstyle B} = \frac{q_e \hbar}{2 M_e}$ is the
Bohr magneton, $M_e$ is the electron mass, $\sigma_z$ is the Pauli
matrix.

We shall solve the full Schr\"odinger equation

\begin{equation}
  H \Psi = E \Psi
\end{equation}

\noindent in two stages. First, we shall obtain an exact solution
of the unperturbed Schrodinger equation

\begin{equation}\label{eq7}
  H_{\scriptscriptstyle 0} \Psi_{\scriptscriptstyle 0} = E_{\scriptscriptstyle 0}
   \Psi_{\scriptscriptstyle 0} ,
\end{equation}

\noindent and then we shall take into account the Zeeman
interaction within the framework of the
 perturbation theory.

It is easy to see that in the case of the unperturbed equation
(\ref{eq7}) with the Hamiltonian (\ref{eq4}), in addition to the
obvious integral of motion

\begin{equation}
  \sigma=\frac{\sigma_x - \sigma_y}{\sqrt {2}}
\end{equation}

\noindent there is also a non-trivial integral of motion

\begin{equation}
  L = L_z + \alpha M_{\rm eff}(x - y)(\sigma_x - \sigma_y)/\hbar,
\end{equation}

\noindent  where $L_z$ is the operator of angular momentum.

We look for the  solutions of Eq. (\ref{eq7}) which are
eigenfunctions of operators $\sigma$ and $L$. Then the required
solutions admit a factorization of the form

\begin{equation}\label{eq10}
  \Psi^{\scriptscriptstyle\pm}_{\scriptscriptstyle 0}(x,y) = \textbf{n}^{\scriptscriptstyle\pm}
  \exp \left[\mp i\sqrt{2}\,\alpha M_{\rm eff} (x + y) / \hbar^2 \right]
  e^{im\phi}u(\rho),\quad m = 0,\pm 1,\pm 2,\ldots,
\end{equation}

\noindent where $m$ is the angular momentum quantum number,
$\displaystyle\textbf{n}^{\scriptscriptstyle\pm}$ are eigenvectors
of the operator $\sigma$:

\begin{equation}
  \mathop{\sigma}\textbf{n}^{\scriptscriptstyle\pm}=\pm\textbf{n}^{\scriptscriptstyle\pm}, \quad
  \textbf{n}^{\scriptscriptstyle\pm}=\frac{1}{\sqrt{2}}
  \begin{pmatrix}
    1  \\
    \pm e^{ - i\pi / 4}
  \end{pmatrix}.
\end{equation}

\noindent Here we use the polar coordinates $\rho, \, \phi$ ($x
= \rho \cos \phi, \, y = \rho \sin \phi$).

Introducing dimensionless quantities

\begin{equation}
  r = \frac{\rho}{\rho_o}, \quad
  e_{\scriptscriptstyle 0} = \frac{2 M_{\rm eff}
  \rho_o^2}{\hbar^2}\,E_{\scriptscriptstyle 0}, \quad
  v = \frac{2 M_{\rm eff}\rho_o^2}{\hbar^2}\,V , \quad
  a = \frac{2 M_{\rm eff}\rho_o}{\hbar^2} \,\alpha , \quad
  b = \frac{q_e \rho_o^2}{2\hbar}\,B,
\end{equation}

\noindent we write the radial equation

\begin{equation}\label{eq13}
  \frac{d^2u}{d\,r^2} + \frac{1}{r} \frac{d\,u}{d\,r} + \left( e_{\scriptscriptstyle 0} + a^2
  - v_c(r) - \frac{m^2}{r^2} - 2\,b\,m - b^2 r^2 \right) u = 0,
\end{equation}

\noindent where

\begin{equation}\label{eq14}
  v_c(r) =
  \begin{cases}
    v, &  0 < r < r_i, \\
    0, &  r_i < r < 1, \\
    v, &  1 < r < \infty.
  \end{cases}
\end{equation}

\noindent Here we use the notation $r_i = \rho_i / \rho_o$ for the relative width of the ring.

In the case of a rectangular potential well (\ref{eq14}) in each
of the three regions ( $0 < r < r_i$, $r_i < r < 1$, $1 < r <
\infty$ ), Eq. (\ref{eq13}) coincides with the equation for the
radial wave function of an electron in a uniform magnetic field
without
 taking into account the spin-orbit interaction \cite{landau}, provided  we replace
 $e_{\scriptscriptstyle 0}$ to $e_{\scriptscriptstyle 0} + a^2 - v_c(r)$. Therefore,
  following \cite{landau}, we represent a required function $u(r)$
  by the formula

\begin{equation}
u(r) = \left(b r^2\right)^{\frac{\lvert m\rvert}{2}}
\exp\left(\frac{-b r^2}{2}\right) w(r),
\end{equation}

\noindent where

\begin{equation}
  w(r) =
  \begin{cases}
    c_1 w_1(r), &  0 < r < r_i, \\
    c_{21} w_{21}(r) + c_{22} w_{22}(r), &  r_i < r < 1, \\
    c_3 w_3(r), &  1 < r < \infty.
  \end{cases}
\end{equation}

\noindent Here, $c_1$, $c_{21}$, $c_{22}$ and $c_3$ are arbitrary
coefficients, and the functions $w_1(r)$, $w_{21}(r)$, $w_{22}(r)$
and $w_3(r)$  are expressed in terms of the confluent
hypergeometric functions of the first and second kind
$M(\alpha,\beta,\xi)$ and $U(\alpha,\beta,\xi)$ \cite{abr} as
follows

\begin{equation}
  \begin{gathered}
    w_1(r) = M\left(\gamma_o,\beta,b r^2\right), \\
    w_{21}(r) = M\left(\gamma_i,\beta,b r^2\right), \quad
    w_{22}(r) = U\left(\gamma_i,\beta,b r^2\right), \\
    w_3(r) = U\left(\gamma_o,\beta,b r^2\right),
  \end{gathered}
\end{equation}

\noindent where

\begin{equation}
  \begin{gathered}
    \gamma_o = \frac{m + \lvert m\rvert + 1}{2}
    - \frac{e_{\scriptscriptstyle 0} + a^2 - v}{4 b}, \\
    \gamma_i = \frac{m + \lvert m\rvert + 1}{2}
     - \frac{e_{\scriptscriptstyle 0} + a^2}{4 b},  \\
    \beta = \lvert m\rvert + 1.
  \end{gathered}
\end{equation}

\noindent The particular solutions in the first and third regions
are chosen so that the radial
 wave function is regular at the origin $r \rightarrow 0$ and tends to zero at the infinity
  $r \rightarrow \infty$.

The continuity conditions for the radial wave function $u(r)$ and its first derivative
 $u^\prime(r) = d u(r)/d r$ at the boundary points $r = r_i$ and $r = 1$ lead to a system
 of algebraic equations

\begin{equation}
  M_4(m,e_{\scriptscriptstyle 0},v,a,b,r_i){\bf X} = 0
\end{equation}

\noindent for four coefficients, where ${\bf X} =
\{c_1,c_{21},c_{22},c_3 \}$ and $M_4(m,e_{\scriptscriptstyle
0},v,a,b,r_i)$ is $4 \times 4$ matrix of the form

\begin{equation}
  M_4 =
  \begin{pmatrix}
    w_1(r_i) & -w_{21}(r_i) & -w_{22}(r_i) & 0 \\
    w_1^\prime(r_i) & -w_{21}^\prime(r_i) & -w_{22}^\prime(r_i) & 0 \\
    0 & w_{21}(1) & w_{22}(1) & -w_3(1) \\
    0 & w_{21}^\prime(1) & w_{22}^\prime(1) &  -w_3^\prime(1)
  \end{pmatrix}.
\end{equation}

\noindent Therefore, the exact equation for determining
$e_{\scriptscriptstyle 0}(m,v,a,b,r_i)$ reeds

\begin{equation}\label{eq25}
  \det M_4(m,e_{\scriptscriptstyle 0},v,a,b,r_i) = 0.
\end{equation}

From Eq. (\ref{eq13}), we see that the dependence of
$e_{\scriptscriptstyle 0}$  on $a$ is trivial $
  e_{\scriptscriptstyle 0}(m,v,a,b,r_i) = e_{\scriptscriptstyle 0}(m,v,0,b,r_i) - a^2$.
 In addition, the following relation $
  e_{\scriptscriptstyle 0}(m,v,a,b,r_i) - e_{\scriptscriptstyle 0}(-m,v,a,b,r_i) = 4b\,m
$  is fulfilled. Of course, Eq. (\ref{eq25}) cannot be solved
analytically, but can be easily solved numerically.

In order to construct the radial wave function completely, we find
the values of required coefficients

\begin{equation}
  \begin{pmatrix}
    c_{21}  \\
    c_{22}  \\
    c_3  \\
  \end{pmatrix}
  = -c_1 M_3^{-1}(m,e_{\scriptscriptstyle 0},v,a,b,r_i)
  \begin{pmatrix}
    w_1^\prime(r_i)  \\
    0  \\
    0  \\
  \end{pmatrix} ,
\end{equation}

\noindent where

\begin{equation}
  M_3 =
  \begin{pmatrix}
    -w_{21}^\prime(r_i) & -w_{22}^\prime(r_i) & 0 \\
    w_{21}(1) & w_{22}(1) & -w_3(1) \\
    w_{21}^\prime(1) & w_{22}^\prime(1) &  -w_3^\prime(1)
  \end{pmatrix}.
\end{equation}

\noindent The residual arbitrariness in the choice of the
coefficient $c_1$ is used to
 implement the standard normalization condition
$
  \langle\Psi^{\scriptscriptstyle\pm}_{\scriptscriptstyle 0}|
  \Psi^{\scriptscriptstyle\pm}_{\scriptscriptstyle 0}\rangle = 1
$.

\section{Contribution of the Zeeman interaction within the
framework of the
 perturbation theory}

The expression for the Zeeman interaction in dimensionless
quantities takes the form

\begin{equation}
  h' = \frac{2 M_{\rm eff} \rho_o^2}{\hbar^2}\,H' = 4 s b\,\sigma_z,
  \quad s = \frac{gM_{\rm eff}}{4M_e}.
\end{equation}

Since each energy level of the unperturbed system is doubly
degenerate with two eigenfunctions (\ref{eq10}), we consider the
contribution of  Zeeman interaction  with the help of the
perturbation theory in the degenerate case.

Because of
$\mathop{\sigma_z}\textbf{n}^{\scriptscriptstyle\pm}=\textbf{n}^{\scriptscriptstyle\mp}$
 in the basis of the eigenvectors $\displaystyle|\Psi^{\scriptscriptstyle +}_{\scriptscriptstyle
0}\rangle$ and $\displaystyle|\Psi^{\scriptscriptstyle
-}_{\scriptscriptstyle 0}\rangle$ of the unperturbed Hamiltonian,
we have the following equalities

\begin{equation}
\displaystyle\langle\Psi^{\scriptscriptstyle\pm}_{\scriptscriptstyle
0}| \sigma_z|\Psi^{\scriptscriptstyle\pm}_{\scriptscriptstyle
0}\rangle = 0 .
\end{equation}

\noindent for the diagonal matrix elements. Off-diagonal matrix
elements are given by

\begin{equation}
  \langle\Psi^{\scriptscriptstyle +}_{\scriptscriptstyle 0}|\sigma_z|\Psi^{\scriptscriptstyle
  -}_{\scriptscriptstyle 0}\rangle =  \langle\Psi^{\scriptscriptstyle -}_{\scriptscriptstyle 0}|\sigma_z|\Psi^{\scriptscriptstyle
  +}_{\scriptscriptstyle 0}\rangle =  \delta(m,v,a,b,r_i),
\end{equation}

\noindent where

\begin{equation}\label{eq32}
  \delta(m,v,a,b,r_i)=\frac{\displaystyle\int_0^\infty J_0(2ar)u^2(r)r\,dr}
  {\displaystyle\int_0^\infty u^2(r)r\,dr}
\end{equation}

\noindent is expressed in terms of the Bessel function.

Then we get splitting

\begin{equation}
  e^{\scriptscriptstyle\pm}=e_{\scriptscriptstyle 0} \pm e'
\end{equation}

\noindent for the unperturbed energy levels, where

\begin{equation}\label{eq32}
  e'\equiv e'(m,v,a,b,r_i)=4s b \delta(m,v,a,b,r_i) .
\end{equation}

\noindent The following relation $
  e'(-m,v,a,b,r_i)=e'(m,v,a,b,r_i)
$ is fulfilled for the corrections $e'$.  Normalized
eigenfunctions in zero-order approximation, which correspond to
the eigenvalues $e^{\scriptscriptstyle\pm}$, are described by the
formulas

\begin{equation}\label{eq34}
  \Psi^{\scriptscriptstyle\pm} = \frac{1}{\sqrt{2}}
  \left(\Psi^{\scriptscriptstyle +}_{\scriptscriptstyle 0} \pm
  \Psi^{\scriptscriptstyle -}_{\scriptscriptstyle 0}\right).
\end{equation}

\noindent Note that in the limiting case $\alpha = 0$ the
expressions (28) and (\ref{eq34}) become exact.

\section{Numerical results}

Now we present  the results of the calculations of the quantities
$e_{\scriptscriptstyle 0}$ and $e'$.
 In accordance with \cite{tsi} we choose the parameter  values
 $M_{\rm eff}=0.067 M_e$, $g=-0.44$ related to GaAs. Then we get $s=-0.00737$.
 If we assume
  $\rho_o=30$~nm, then
 the following correspondences $a = 1 \to \alpha = 18.9579$~meV nm, $e = 1\to
 E = 0.631933$~meV between the dimensionless and dimensional quantities are obtained.
  For example, at
  chosen parameters, the dimensionless value  $v =400$
corresponds to the potential well  depth $V = 252.772$~meV, which
is close to the value $257$~meV  in \cite{gro}.
\begin{figure} [h]
\centering
\includegraphics[width=16cm]{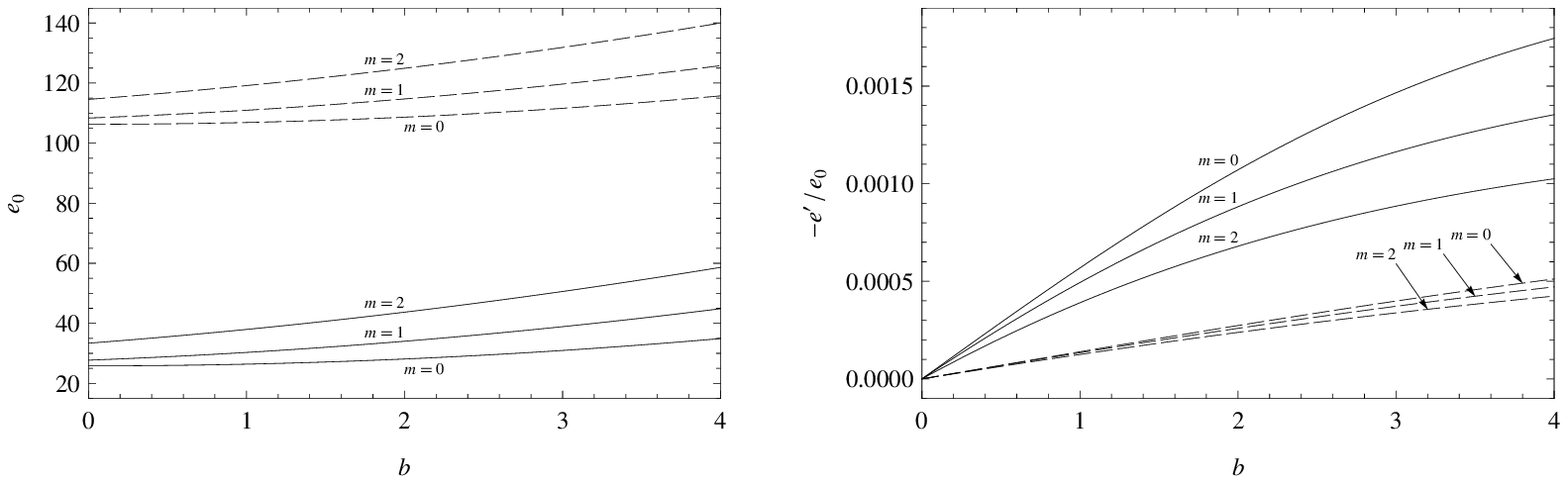}\\
\caption{Dependence of  $e_{\scriptscriptstyle 0}$ and
$-e'/e_{\scriptscriptstyle 0}$ on $b$  at  $a = 1$ and $r_i =
0.5$.}
\end{figure}
\begin{figure}[h]
\centering
\includegraphics[width=16cm]{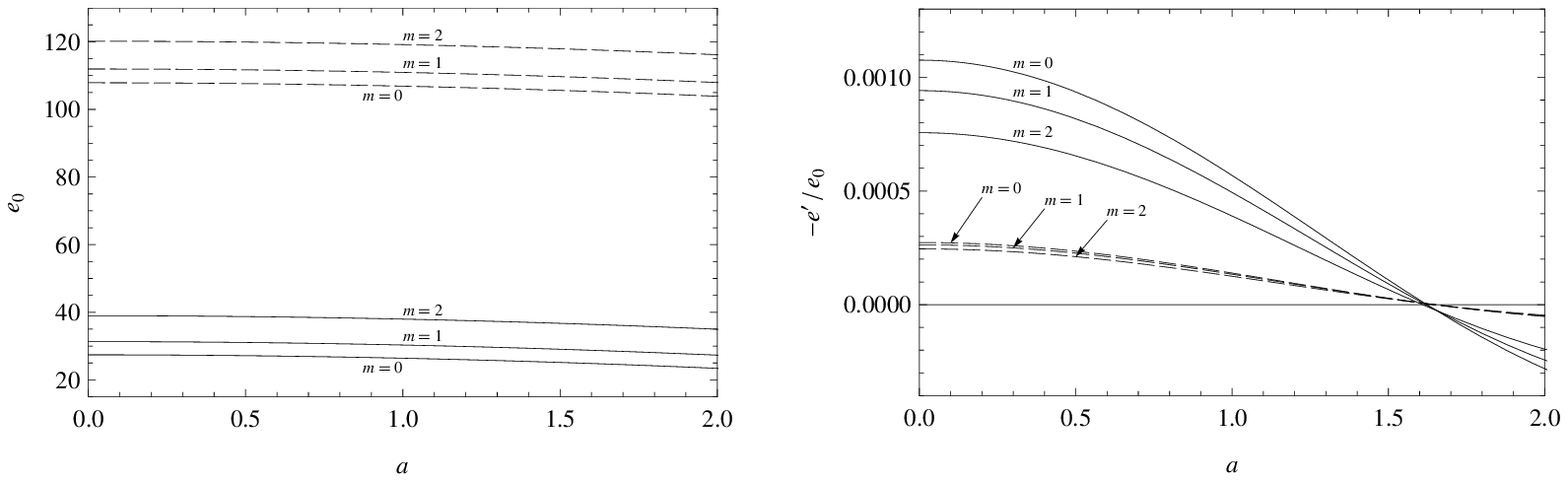}\\
\caption{Dependence of  $e_{\scriptscriptstyle 0}$ and
$-e'/e_{\scriptscriptstyle 0}$ on $a$  at  $b = 1$ è $r_i = 0.5$.
}
\end{figure}

Fig. 1 shows in dimensionless units  the dependence of the
unperturbed energy $e_{\scriptscriptstyle 0}$ and the relative
corrections $-e'/e_{\scriptscriptstyle 0}$ to the energy on a
magnetic field $b$ for the fixed values  $a = 1$ and $r_i = 0.5$.
Fig. 2  demonstrates the dependence of $e_{\scriptscriptstyle 0}$
and
 $-e'/e_{\scriptscriptstyle 0}$ on the spin-orbit interaction  strength $a$ for fixed
 values  $b = 1$ and $r_i = 0.5$.
 The solid lines represent the first energy levels and the dashed
lines represent the second levels for three values of the angular
momentum ($m=0,1,2$).

\begin{table}[h]
\caption{ Dependence of  $e_{\scriptscriptstyle 0}$  and
$-e'/e_{\scriptscriptstyle 0}$ on $r_i$ at $a=1$ and $b=1$.}
\begin{center}
{\scriptsize
\begin{tabular}{r|*6c}
 \hline
$r_i$ & \multicolumn{2}{c}{$m=0$} & \multicolumn{2}{c}{$m=1$} & \multicolumn{2}{c}{$m=2$} \bigstrut \\
 \cline{2-7}
                                        & \multicolumn{6}{c}{$e_{\scriptscriptstyle 0}, \quad (-e'/e_{\scriptscriptstyle 0})$} \bigstrut \\
  \hline
0                                        & 4.48334 & 26.952 & 14.677 & 45.9308 & 27.3538 & 67.4988 \bigstrut[t]\\
                                         & $(5.1620\times 10^{-3})$ & $(7.6793\times 10^{-4})$ & $(1.3573\times 10^{-3})$ & $(4.3783\times 10^{-4})$ & $(6.5366\times 10^{-4})$ & $(2.8612\times 10^{-4})$ \bigstrut[b]\\
0.1                                      & 7.72294 & 36.3199 & 14.9000 & 47.1566 & 27.3612 & 67.5757 \bigstrut[t]\\
                                         & $(2.7373\times 10^{-3})$ & $(5.5037\times 10^{-4})$ & $(1.3272\times 10^{-3})$ & $(4.2162\times 10^{-4})$ & $(6.5331\times 10^{-4})$ & $(2.8550\times 10^{-4})$ \bigstrut[b]\\
0.5                                      & 26.4059 & 106.878 & 30.3106 & 110.949 & 37.9733 & 119.165 \bigstrut[t]\\
                                         & $(5.6869\times 10^{-4})$ & $(1.3905\times 10^{-4})$ & $(4.9374\times 10^{-4})$ & $(1.3401\times 10^{-4})$ & $(3.9012\times 10^{-4})$ & $(1.2497\times 10^{-4})$ \bigstrut[b]\\
0.9                                      & 218.124 & 404.788 & 221.245 & 407.21 & 226.608 & 410.102 \bigstrut[t]\\
                                         & $(3.8120\times 10^{-5})$ & $(-1.2162\times 10^{-5})$ & $(3.7573\times 10^{-5})$ & $(-2.0380\times 10^{-5})$ & $(3.6655\times 10^{-5})$ & $(-1.9963\times 10^{-5})$ \bigstrut[b]\\
\hline
\end{tabular}}
\end{center}
\end{table}

Table 1. illustrates the significant dependence of the unperturbed
energy $e_{\scriptscriptstyle 0}$ and the relative correction
$-e'/e_{\scriptscriptstyle 0}$  on the relative width $r_i$ of the
quantum ring at the fixed values  $a = 1$ and $b = 1$. The values
of $-e'/e_{\scriptscriptstyle 0}$ are placed in parentheses. The
value $r_i = 0$ corresponds to a quantum dot. The table lists only
 first two energy levels. Note that for $r_i = 0.9$ already the
second energy levels lie above the potential  well with the
dimensionless depth of $v = 400$ in the considered case. In the
absence of a magnetic field, the bound states occur only at $
e_{\scriptscriptstyle 0} < v$. The presence of a magnetic field
leads to appearance of the discrete energy levels exceeding
magnitude of a potential well depth.

\section{Conclusion}

Wave functions and energy levels are obtained for  electrons in
two-dimensional quantum rings with taking into account the
spin-orbit Rashba and Dresselhaus interactions of equal strengths
in the presence of an external uniform constant magnetic field in
the framework of an adequate model with a confining potential of
finite depth. These results  may be of interest in the study of
spin-dependent effects in semiconductor heterostructures.

\end{document}